# Crystal structure and superconducting properties of hexagonal lithium-niobium oxynitride

Teruki Motohashi,[*,†] Masahiko Ito,[†] Yuji Masubuchi,[†] Makoto Wakeshima,[‡]
and Shinichi Kikkawa[†]

[†]*Faculty of Engineering, Hokkaido University, Sapporo 060-8628, Japan, and*
[‡]*Faculty of Science, Hokkaido University, Sapporo 060-0810, Japan*

A hexagonal oxynitride $(Li_{0.88}\square_{0.12})Nb_{3.0}(O_{0.13}N_{0.87})_4$ was synthesized through ammonia nitridation of $LiNb_3O_8$. The structural analysis revealed that this oxynitride consists of alternate stacking of octahedral and prismatic layers with different Li/Nb ratios: significant amounts of Li and Nb atoms (Li/Nb = 43/57) coexist in the octahedral layer, while the prismatic site is preferentially occupied by Nb (Li/Nb = 3/97). A metallic behavior was accompanied by an abrupt drop of electrical resistivity at about 3 K. Furthermore, large diamagnetism and specific-heat anomaly were observed below this temperature, suggesting the appearance of superconductivity in the Li-Nb oxynitride.

*Corresponding author. Faculty of Engineering, Hokkaido University, N13, W8, Kita-ku, Sapporo 060-8628, Japan. Tel: +81(0)11 706 6741. Fax: +81(0)11 706 6740. E-mail: t-mot@eng.hokudai.ac.jp

## INTRODUCTION

Since the discovery of high temperature superconductivity in layered copper oxides, there has been great interest in layered compounds as promising candidates for novel superconductors. It is widely believed that the appearance of superconductivity in copper oxides is attributed to their two-dimensional (2D) electronic structure, which is subject to electronic instabilities of spin/charge long-range order, leading to unconventional electron pairing mechanisms.[1] Layered compounds are also fascinating to realize high critical temperatures ($T_c$) within the conventional BCS framework,[2] as the low dimensionality favors a larger density of states at the Fermi level. In fact, recent research on the exploration of layered compounds led to the discoveries of superconductors with high $T_c$ values, such as layered hafnium nitride chlorides ($T_c$ ~ 25.5 K)[3] and layered iron arsenides ($T_c$ ~ 55 K).[4] It should be noted that some layered superconductors were highlighted because of their unconventional superconductivity, although their $T_c$ values are low: e.g., 1.4 K of $Sr_2RuO_4$ (Ref. 5) and 3.5 K of $Na_xCoO_2$ $y H_2O$.[6]

Niobium nitrides/oxynitrides are known as potential materials for studies on superconductivity. They crystallize in various types of rock-salt-derived structures depending on the chemical composition and synthesis condition: e.g., cubic δ-NbN, tetragonal $Nb_4N_5$, and hexagonal $Nb_5N_6$. While most of the Nb nitrides/oxynitrides are superconductors with relatively high $T_c$ values, e.g., $T_c$ = 14.7 ~ 17.7 K for δ-NbN,[7-9] $T_c$ = 9.6 ~ 14.7 K for δ-Nb(O,N),[10] and $T_c$ = 8.5 ~ 10 K for $Nb_4N_5$,[11-13] the hexagonal $Nb_5N_6$ phase was reported to be nonsuperconductive despite the similarity in its Nb/N ratio.[12] Noticeably, the hexagonal $Nb_5N_6$ phase has a characteristic structural feature; the crystal of this phase consists of an AABB arrangement of nitrogen atoms, with the stacking sequence AαAγBβBγ where α and β denote trigonal prismatic sites and γ octahedral sites.[11] The arrangement of niobium atoms is assumed to be identical to that of tantalum atoms in hexagonal $Ta_5N_6$, in which cationic deficiencies exist only in the octahedral layer as shown in Fig. 1(a). Hence, the chemical formula of this nitride can be written as $(Nb_4\square_2)^{octahedron}(Nb_6)^{prism}N_{12}$. This feature is in contrast to that of cubic δ-NbN, which contains an ABCABC arrangement of nitrogen atoms with a three-dimensional (3D) network of octahedral sites only.

We focused on complex nitrides/oxynitrides with a $Nb_5N_6$-related hexagonal structure, as the distinct coordination environments in the prismatic and octahedral layers may induce site preferences of the constituent cations to form layered structures. It is remarkable to clarify whether their layered atomic arrangements are capable for the appearance of superconductivity. F. Tessier et al. reported that hexagonal $LiNb_3N_{3.7}$ is formed through ammonia nitridation of the ternary oxide $LiNb_3O_8$ at 1000 °C.[13-15] They suggested that the product was essentially a nitride containing a negligible amount of oxygen. The crystal structure is isostructural to $LiTa_3N_4$ which consists of a layered arrangement of trigonal prismatic Ta sites and octahedral Li/Ta sites, with a general formula represented by $(Li_3Ta_3)^{octahedron}(Ta_6)^{prism}N_{12}$.[16] This Li-Nb oxynitride may be an anisotropic electron conductor because of the 2D prismatic Nb layer sandwiched by two octahedral Li/Nb layers. Nevertheless, details of its crystal structure and electrical properties were not reported.

In the present work, structural and electrical characteristics were investigated for the hexagonal Li-Nb oxynitride. Its crystal structure was refined with neutron diffraction at room temperature. The electronic properties were measured down to 0.45 K. Our electrical resistivity measurement revealed a metallic behavior, accompanied by an abrupt resistivity drop at about 3 K. A large diamagnetism and specific-heat anomaly were observed below this temperature, suggesting the appearance of superconductivity in a layered oxynitride lattice.

## EXPERIMENTAL SECTION

Polycrystalline samples of the Li-Nb oxynitride were prepared by ammonia nitridation of $LiNb_3O_8$ as an oxide precursor, as reported in the literature.[13-15] A mixture of $Li_2CO_3$ (Kanto Chemicals, 99%; dried at 110°C prior to use) and $Nb_2O_5$ (Wako Pure Chemicals, 99.9%) powders with the ratio of Li/Nb = 1/3 was fired at 1000°C for 24 h in air to synthesize pure $LiNb_3O_8$.[17] Approximately 250 mg of the resultant $LiNb_3O_8$ powder was pelletized and then nitrided in flowing ammonia (Sumitomo Seika Chemicals, 99.9%) of 200 mL/min at 1000°C for 15 h, followed by slow cooling to room temperature. Mass products (~ 1.5 g) for neutron diffraction were also obtained through ammonia nitridation of the $LiNb_3O_8$ powder at 1000°C for 10 h.

Li and Nb compositions of the products were analyzed by inductively coupled plasma-atomic emission spectroscopy (ICP-AES; Shimadzu ICPE-9000). A 10 mg of the sample powder was successfully dissolved in a mixed aqueous solution of 0.01 M ammonium oxalate (90 mL) and 30% hydrogen peroxide (10 mL). The oxygen and nitrogen contents were determined using a combustion analyzer (EMGA-620, Horiba), which was calibrated with $Y_2O_3$ (Wako Pure Chemicals, 99.99%; fired at 1000°C for 10 h prior to use) and $Si_3N_4$ (The Ceramic Society of Japan, JCRM R 003) as standard materials.

Phase purity and lattice parameters of the products were checked by means of X-ray diffraction (XRD; Rigaku Ultima IV; Cu Kα radiation). The data were collected over the 2θ angular range of 5 ~ 90° with a step size of 0.02°. Neutron diffraction (ND) measurements were also performed to refine the crystal structure of the Li-Nb oxynitride. A high resolution powder diffractometer (HRPD; HERMES, Institute for Materials Research, Tohoku University) installed at the JRR-3M reactor of the Japan Atomic Energy Agency (JAEA) was utilized for the data acquisition at room temperature. A neutron beam with a wavelength of 0.184780(15) nm was obtained with the 331 reflection of a Ge monochromator. ND was measured at 2θ = 3.0 ~ 152.9° with a step size of 0.1° using approximately 6 g of the sample powder filled in a vanadium cylinder (10 mmϕ). The crystal structure of the Li-Nb oxynitride was refined by the Rietveld method using RIETAN-FP program[18] and visualized with VESTA software.[19]

The bulk product was cut into rectangular pieces with a typical size of 3.5 × 7.0 × 1.0 $mm^3$ for physical property measurements. Electrical resistivity (ρ) was measured employing a four-probe technique (Quantum Design, PPMS) with an applied current of 1 mA in a temperature range of 2 ~ 300 K. Magnetic susceptibility (χ) was measured using a superconducting quantum interference device magnetometer (SQUID; Quantum

Design, MPMS-S) in a temperature range of 1.7 ~ 300 K under a magnetic field of 5 Oe. Specific heat ($C_p$) measurements were performed using a relaxation technique with a commercial apparatus (Quantum Design, PPMS) equipped with a $^3$He/$^4$He dilution refrigerator between 0.45 and 25 K. A 15 mg bulk sample was mounted on a thin alumina plate with Apiezon N-grease for better thermal contact.

## RESULTS

**Structural characterization.** A single phase product of the hexagonal Li-Nb oxynitride was synthesized, as evidenced by a powder X-ray diffraction pattern in Fig. 2. The product was black in color with slight luster at the grain surface, suggestive of their metallic behavior. The result of chemical analyses (Table 1) indicates that the product is cation-deficient with the Li/Nb ratio being slightly smaller than the starting ratio. It is also worth noting that a certain amount of oxygen was detected by the combustion analysis. The product was well crystallized consisting of relatively large grains with ~ 1μm in diameter; hence the concern that the chemical analysis might suffer from the extrinsic oxidic surface contamination of nitride grains can be ruled out. We thus conclude that the product is not a nitride but an oxynitride: the larger amount of oxygen in our product than that of Tessier et al.[14,15] is likely attributed to the smaller ammonia flow rate in the present work. The lattice parameters of the hexagonal phase are slightly large for our product ($a$ = 0.52043 nm and $c$ = 1.03942 nm) than those reported in Refs. 13-15 ($a$ = 0.52023 nm and $c$ = 1.0363 nm), probably because of the larger amount of oxygen.

The crystal structure of the hexagonal Li-Nb oxynitride was refined. As a first step, the distribution of heavy Nb atoms in the unit cell was determined by the Rietveld fit of the XRD data. The calculation was made on the basis of the $LiTa_3N_4$-type structure (hexagonal space group $P6_3/mcm$) that contains three cationic sites, i.e., the octahedral 2*b*, 4*d*, and prismatic 6*g* sites. In this analysis, we assumed for simplicity that the contribution from Li atoms was neglected, and oxygen and nitrogen were located statistically at a single anionic site. The Nb occupancies were converged to be 0.29(1), 0.65(1), and 0.97(1) for the 2*b*, 4*d*, and 6*g* sites, respectively.

Next, all the atomic parameters were refined with the ND data. The neutron scattering length of Li is relatively large with a negative sign such that determination of the Li occupancies may be possible. The values used in the refinement were -1.90 for Li, 7.054 for Nb, 5.803 for O, and 9.36 for N.[18] The powder used for neutron diffraction contained a trace amount of low-temperature Li-Nb oxynitride phase (cubic *Fm*-3*m*)[13-15] and $Li_3NbO_4$ (cubic *I*-4$_3$*m*)[20] as secondary phases (2 ~3 wt% for each), which were formed because of inevitable inhomogeneity in the mass product. These secondary phases were also included in the refinement. The chemical composition of the cubic Li-Nb oxynitride impurity was fixed at $(Li_{0.20}Nb_{0.71}\square_{0.09})(O_{0.22}N_{0.78})$, as determined for the product nitrided at 800°C (see Supporting Information), since the exact composition was unknown. The displacement factors (*B*) at the cationic 2*b*, 4*d*, and 6*g* sites were refined setting to be equal. The final convergence calculation without compositional constraints for all the elements led to an atomic ratio which is in perfect agreement with the result of chemical analysis. The refined chemical formula is

$(Li_{0.88}\square_{0.12})Nb_{3.0}(O_{0.13}N_{0.87})_4$.

Figure 3 presents the ND data together with the fit result of the Li-Nb oxynitride, and the resultant crystal structure is visualized in Fig. 1(b). Atomic parameters and refinement details are summarized in Tables 2 and 3. It is obvious that the Nb atoms preferentially occupy the prismatic 6*g* site, as suggested by the previous work.[13-15] Another structural model was tested locating 100% Nb atoms (i.e., no Li atoms) at the 6*g* site, but the goodness of fit apparently deteriorated. This suggests that a small part of Nb atoms at the 6*g* site is substituted by Li. Meanwhile, significant amounts of Li and Nb atoms coexist at the octahedral 2*b* and 4*d* sites. Interestingly, the 2*b* site contains approximately 2/3 Li and 1/3 Nb, while the 4*d* site with an opposite Li/Nb ratio. Such cationic distribution results in nonequivalent local environments at these sites (see Table 3), that arise from the anionic site slightly deviated from the ideal position at $x$ = 1/3. The absence of super structure peaks in the X-ray and neutron diffraction patterns implies that oxygen/nitrogen atoms are statistically distributed at the single site. Meanwhile, in some oxynitrides such as $SrTaO_2N$ there is a preference for slightly different sites for O and N.[21-23] The existence of possible short-range O/N order in this Li-Nb oxynitride cannot be ruled out and needs to be investigated utilizing, e.g., selected area electron diffraction.

**Physical properties.** Electrical resistivity (ρ) of the hexagonal Li-Nb oxynitride is plotted as a function of temperature in Fig. 4. The magnitude of ρ is approximately $10^{-2}$ Ω m at 300 K, and almost independent of temperature down to 4 K. Such a highly conductive behavior without carrier localization strongly indicates that this oxynitride is essentially a metal, in spite of a small negative temperature coefficient ($d\rho/dT < 0$) which would originate from weak links at grain boundaries. Upon further cooling, the magnitude of $\rho$ starts to decrease abruptly at about 3 K, as highlighted in the inset of Fig. 4. The $\rho$ value at 2 K is $5.0 \times 10^{-4}$ $\Omega$ m: this value is by a factor of 1/24 smaller than that at 4 K. Unfortunately, measurements below 2 K were not possible because of the instrumental limitation.

In the magnetic susceptibility ($\chi$) measurement, a diamagnetic signal was observed below 2.7 K both in the zero-field-cooling (ZFC) and field-cooling (FC) curves under a low field (5 Oe), as shown in Fig. 5. The onset of diamagnetism is in good agreement with the temperature where the resistivity drops. The magnitude of $\chi$ reaches $-1.20 \times 10^{-3}$ emu $g^{-1}$ $Oe^{-1}$ at 1.7 K in the ZFC curve: this value corresponds to 11% superconducting volume fraction ($V_{SC}$), assuming that the diamagnetism comes from superconductivity in the Li-Nb oxynitride. It should be noted that the diamagnetic signal gradually increases with lowering temperature below 2.7 K. This implies that the sample involves electronic inhomogeneity which is likely to originate from local differences in the Li/Nb ratio. Note that the neutron diffraction only tells the averaged atomic distribution. It is thus likely that the magnitude of $V_{SC}$ could be enhanced upon further cooling below 1.7 K.

The result of the specific heat measurement is presented in a $C_p\ T^{-1}$ vs $T^2$ plot (Fig. 6). The data points fall on a straight line above 4 K with a finite intercept at the vertical axis, typically seen for metallic compounds with significant amounts of carriers, consistent with our interpretation of the resistivity data. At low temperatures, the plot deviates from the straight line, showing a hump around 1 ~ 3 K and then a convergence to zero when $T \to 0$ K. This convergence strongly indicates a gap opening at the Fermi level, i.e., the appearance of superconductivity. For a typical superconductor, a discontinuous $C_p$ jump is observed at $T_c$. We interpret the hump in our data as a gradual $C_p$ jump due to the local differences in $T_c$. This aspect is consistent with the magnetic susceptibility data.

## DISCUSSION

The present work revealed the appearance of superconductivity at about 3 K in the hexagonal Li-Nb oxynitride products. From the fact that the actual product characterized by resistivity, magnetization, and specific heat does not suffer from contamination of any secondary phases (Fig. 2), it is evident that the superconducting response definitely originates from the hexagonal phase. Note that the coexistence of a Li-free superconductive phase is unlikely, since hexagonal $Nb_5N_6$ was reported to be nonsuperconductive[12], as mentioned in the Introduction section.

The hexagonal Li-Nb oxynitride is featured with its layered crystal structure which consists of alternate stacking of octahedral and prismatic layers with different Li/Nb ratios, represented as $(Li_{2.4}Nb_{3.2}\square_{0.4})^{octahedron}(Li_{0.2}Nb_{5.8})^{prism}(O_{0.13}N_{0.87})_{12}$. Taking into account such a structural feature, the electronic structure is likely to be anisotropic. In fact, electrical properties of the Li-Nb oxynitrides are found to be correlated with the atomic arrangement. The cubic Li-Nb oxynitride with a similar chemical composition, $(Li_{0.20}Nb_{0.71}\square_{0.09})(O_{0.22}N_{0.78})$, was obtained by ammonia nitridation of $LiNb_3O_8$ at 800°C. The crystal structure of this cubic phase was refined to be rock-salt type where a single cationic site is statistically occupied by Li and Nb (see Supporting Information for details). The cubic phase showed a strong semiconductive upturn without any signature of superconductivity down to 2 K, as shown in Fig. 7. This result implies that the existence of 20% Li atoms at the Nb site is sufficient to disrupt the metallic conduction path. Thus, the metallic behavior in the hexagonal phase could be attributed to the prismatic layer, while the Li-rich octahedral layer acts as an electrical blocking block along the stacking direction, as in the layered copper oxide superconductors.

From the result of the chemical analysis, the oxidation number of niobium is readily calculated at +3.53, which is apparently deviated from the preferred oxidation states of Nb, i.e., III and V. Such an usual oxidation number may be explained assuming the coexistence of two niobium species with distinct coordination environments. One can anticipate that lower/higher-valent Nb ions are located at the prismatic/octahedral sites with longer/shorter interatomic distances, respectively. This interpretation is consistent with the suggestion by Marchand et al.,[13] who assigned $Nb^{3+}$ to prismatic coordination while $Nb^{5+}$ at the octahedral sites.

Within the prismatic layer, the Nb atoms form a 2D triangular lattice with a Nb-Nb distance of ca. 0.3 nm. This value is slightly longer than 0.286 nm for metallic Nb with a body-centered cubic structure.[24] The coordination geometry is similar to those in the known Nb-based superconductors such as $NbSe_2$ ($T_c$ = 7 K)[25] and $Li_xNbO_2$ ($T_c$ = 5 K),[26] but the nature of chemical bonds are apparently different among these compounds.

Interestingly, the triangular Nb lattice involves a small distortion from the regular triangle (see Table 2) so as to form $Nb_3$ trimers, i.e., two shorter distances of 0.298 nm and four longer distances of 0.302 nm. A similar feature is also reported for the isomorphous Li-Ta nitride $Li_{1-x}Ta_{3+x}N_4$ (Ref. 16) and Mn-Nb oxynitride $Mn_{0.54}Nb_{3.07}N_{4.40}O_{0.60}$.[27] It should be noted that such a modulated lattice is also observed in superconductive Ta/Nb dichalcogenides upon charge-density-wave (CDW) formation:[25,28] the interplay between superconductivity and CDW has attracted interest in the physics community.[29] The electronic structure of the prismatic layer, including the influence of the trimer formations, is still unknown and merits further theoretical research.

To gain deeper insight into the superconducting properties of the hexagonal Li-Nb oxynitride, the electronic contribution to the specific heat was estimated. From the linear fit shown in Fig. 6, the electronic specific heat $\gamma$ and Debye temperature $\Theta_D$ were accordingly obtained to be 5.09 mJ $mol^{-1}$ $K^{-2}$ [per $(Li_{0.88}\square_{0.12})Nb_{3.0}(O_{0.13}N_{0.87})_4$ formula unit] and 451 K, respectively. The extracted electronic part of the specific heat ($C_e$) gives a much smaller value of $\Delta C_e/\gamma T_c$ ~ 0.15 than the usual BCS value (1.43). Nevertheless, it should be noted that the $\Delta C_e/\gamma T_c$ value has been severely deteriorated by the broadened nature of the superconducting transition. This implies that a relatively large portion of the product is superconducting at $T \rightarrow 0$ K.

Since the atomic arrangement in the hexagonal Li-Nb oxynitride is largely different from that in the known Nb-based nitride/oxynitride superconductors, it is informative to make comparison of the electronic properties of these compounds. The $T_c$ value of the hexagonal Li-Nb oxynitride is much lower than $T_c$ = 14.7 ~ 17.7 K for the cubic δ-NbN phase.[7-9] In the BCS framework, the magnitude of electronic specific heat γ is one of the key parameters for determining $T_c$, as the γ value is proportional to the density of states at the Fermi level $D(\varepsilon_F)$.[2] The γ value of the hexagonal Li-Nb oxynitride, 5.09 mJ mol$^{-1}$ K$^{-2}$ [per $(Li_{0.88}\square_{0.12})Nb_{3.0}(O_{0.13}N_{0.87})_4$ formula unit], should be divided by 2, when focusing only on the prismatic layer (this consideration has implicitly assumed that the octahedral layer does not contribute to γ or $D(\varepsilon_F)$, although the validity of the assumption is open to dispute). The converted γ value, 2.55 mJ (mol-Nb$^{prism}$)$^{-1}$ K$^{-2}$, is almost identical to that reported for cubic δ-NbN (2.64 mJ mol$^{-1}$ K$^{-2}$).[30]

The coincidence of γ suggests that the $T_c$ value of the hexagonal Li-Nb oxynitride seems to be much lower than anticipated. One of the possible reasons for the suppressed $T_c$ is that the sample suffers from atomic defects within the conductive prismatic layer, as revealed by the ND experiment: i.e., the Li atoms at the prismatic site as scattering centers of carriers could cause electronic inhomogeneity and suppress the maximum $T_c$ value. Nevertheless, the $T_c$ suppression in this oxynitride is too large to be explained only by chemical disorder. It should be emphasized that the $Nb_5N_6$ phase with a related hexagonal structure was reported to be nonsuperconductive. Although the reason for the absence of superconductivity in $Nb_5N_6$ has still been unclear, the electronic structure of the prismatic layer is likely distinct from that within the octahedral layer. In any case, the Li-Nb oxynitride may be a new structural type of nitride/oxynitride superconductors, rather than a member of the conventional rock-salt-type family. Detailed research is highly desirable to investigate the anisotropy in the superconducting properties of this oxynitride.

## CONCLUSIONS

The present work reported the structural and superconducting properties of a hexagonal Li-Nb oxynitride $(Li_{0.88}\square_{0.12})Nb_{3.0}(O_{0.13}N_{0.87})_4$. This oxynitride has a layered structure with alternate stacking of octahedral and prismatic layers. While the octahedral layer contains significant amounts of Li and Nb atoms, the prismatic site is preferentially occupied by Nb, which is likely attributed to the metallic behavior. The results of the physical property measurements strongly suggested that this oxynitride is superconductive with $T_c \approx 3$ K.

**Acknowledgment.** We are grateful to Professor T. Tohyama (Kyoto University) for his comments on the superconducting properties. Also, Professor Y. Hinatsu (Hokkaido University) is acknowledged for his help in the physical property measurements. The present work was supported by Grants-in-aid for Science Research (Contracts No. 21245047 and No. 24654094) from the Japan Society for the Promotion of Science. T.M. acknowledges financial support from the Global COE Program (Project No. B01: “Catalysis as the Basis for Innovation in Materials Science”) from the Ministry of Education, Culture, Sports, Science and Technology, Japan. The neutron diffraction experiment was performed under the approval of 10750.

**Supporting Information.** Structural data of the hexagonal Li-Nb oxynitride phase

(in CIF format), synthesis and crystal structure of the cubic Li-Nb oxynitride phase. This material is available free of charge via the Internet at http://pubs.acs.org.

REREFENCES

Table 1. Chemical composition of the hexagonal Li-Nb oxynitride determined by the ICP-AES and oxygen/nitrogen combustion analyses.

| | Li | Nb | N | O | Total |
|---|---|---|---|---|---|
| wt % | 1.8 ± 0.2 | 82 ± 4 | 14.0 ± 0.3 | 2.3 ± 0.1 | 100.1 ± 4.6 |
| molar ratio | 22 | 75 | 87 | 13 | – |

Table 2. Atomic parameters for the hexagonal Li-Nb oxynitride.

| atom | site | *g* | *x* | *y* | *z* | $B/10^{-2}$ nm$^2$ |
|---|---|---|---|---|---|---|
| Nb | 2*b* | 0.29(1) | 0 | 0 | 0 | 0.06(5) |
| Li | | 0.70(5) | | | | |
| Nb | 4*d* | 0.65(1) | 1/3 | 2/3 | 0 | |
| Li | | 0.26(3) | | | | |
| Nb | 6*g* | 0.97(1) | 0.3306(15) | 0 | 1/4 | |
| Li | | 0.03(2) | | | | |
| N | 12*k* | 0.87(1) | 0.3349(6) | 0 | 0.6214(10) | 0.23(3) |
| O | | 0.13(1) | | | | |

Table 3. Structural refinement details for the hexagonal Li-Nb oxynitride.

| | $(Li_{0.88}\square_{0.12})Nb_{3.0}(O_{0.13}N_{0.87})_4$ |
|---|---|
| crystal system | hexagonal |
| space group | $P6_3/mcm$ |
| $a$ / nm | 0.52043(3) |
| $c$ / nm | 1.03942(5) |
| volume / $nm^3$ | 0.243802(22) |
| $Z$ | 3 |
| calculated density / $10^3$ kg $m^{-3}$ | 7.025 |
| $d_{Li/Nb(2b)\text{-}O/N}$ / nm | (6×) 0.2151 |
| $d_{Li/Nb(4d)\text{-}O/N}$ / nm | (6×) 0.2142 |
| $d_{Li/Nb(6g)\text{-}O/N}$ / nm | (4×) 0.2188 |
| | (2×) 0.2194 |
| $d_{Li/Nb(6g)\text{-}Li/Nb(6g)}$ / nm | (2×) 0.2982 |
| | (4×) 0.3016 |
| Secondary phases | $(Li_{0.20}Nb_{0.71}\square_{0.09})(O_{0.22}N_{0.78})$ (3.0 wt%) |
| | $Li_3NbO_4$ (1.6 wt%) |
| $R_B$ | 1.45 % |
| $R_F$ | 0.95% |
| $R_{wp}$ | 7.94 % |
| $S$ | 2.13 |

## Figure captions

Fig. 1.

Schematic illustrations of the crystal structure of (a) $Nb_5N_6$ and (b) the hexagonal Li-Nb oxynitride. The illustrations were drawn with VESTA software[19] based on the structural model in Ref. 11 for $Nb_5N_6$, and our result of structural refinement for the hexagonal Li-Nb oxynitride.

Fig. 2.

X-ray powder diffraction pattern for the Li-Nb oxynitride bulk used for the physical property measurements. In this pattern, the relative intensity was plotted on a logarithmic scale. The red ticks show the peak positions expected for the hexagonal $P6_3/mcm$ model.

Fig. 3.

Rietveld refinement of the ND data for the hexagonal Li-Nb oxynitride powder.

Fig. 4.

Electrical resistivity (ρ) of the hexagonal Li-Nb oxynitride as a function temperature. The inset shows a magnified ρ - *T* curve below 6 K.

Fig. 5.

Magnetic susceptibility (χ) vs. temperature curve under a magnetic field of 5 Oe for the hexagonal Li-Nb oxynitride.

Fig. 6.

Specific heat ($C_p$) data of the hexagonal Li-Nb oxynitride in a $C_p T^{-1}$ vs. $T^2$ plot.

Fig. 7.

Electrical resistivity (ρ) vs. temperature plots for two types of the Li-Nb oxynitride phases. The black and red symbols denote data of the cubic and hexagonal phases, respectively.

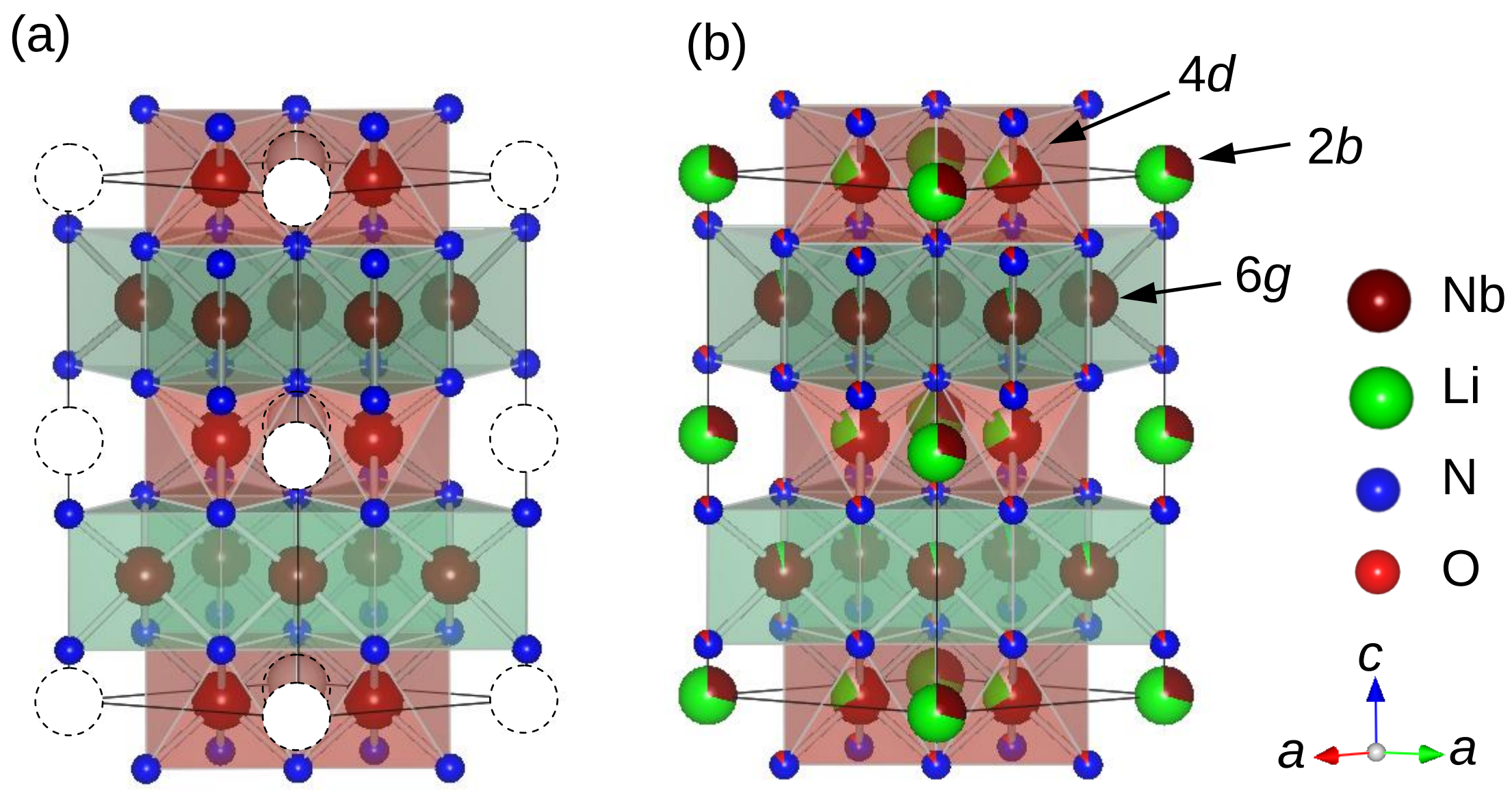

Fig. 1. Motohashi *et al.*

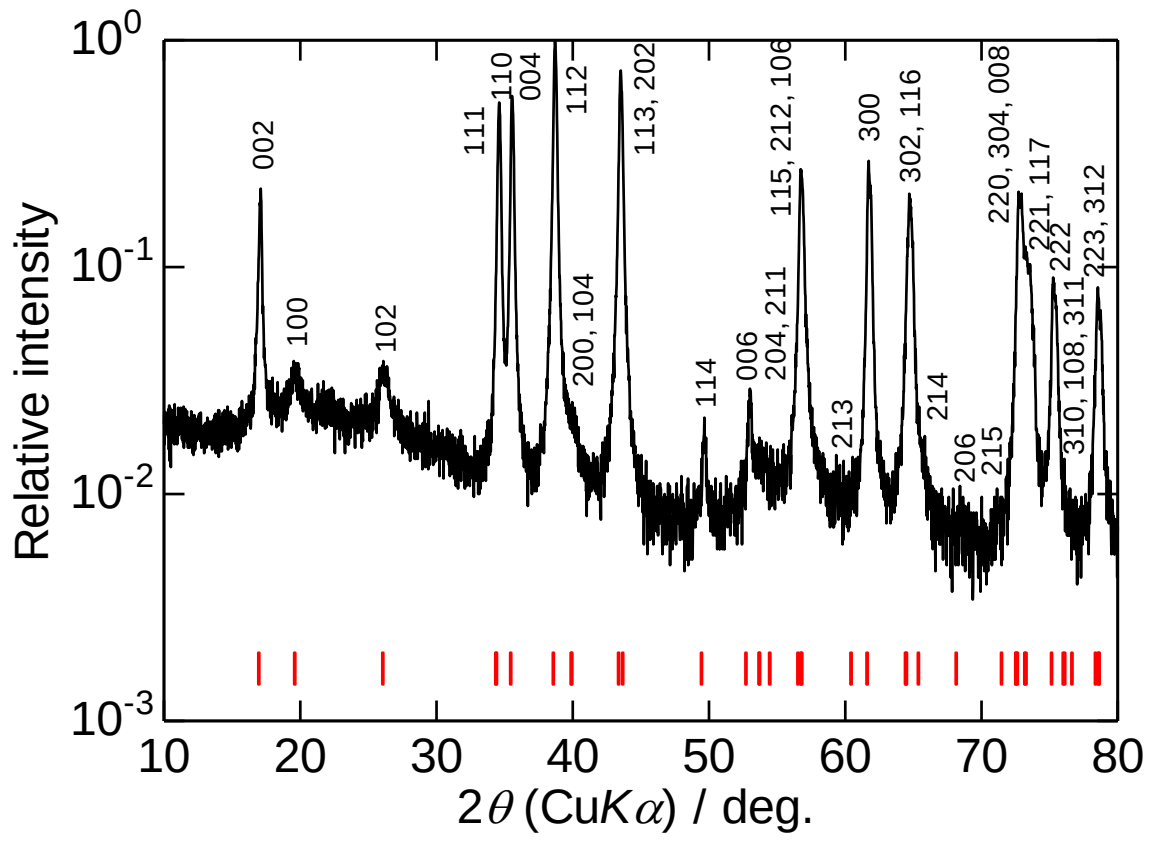

Fig. 2. Motohashi *et al.*

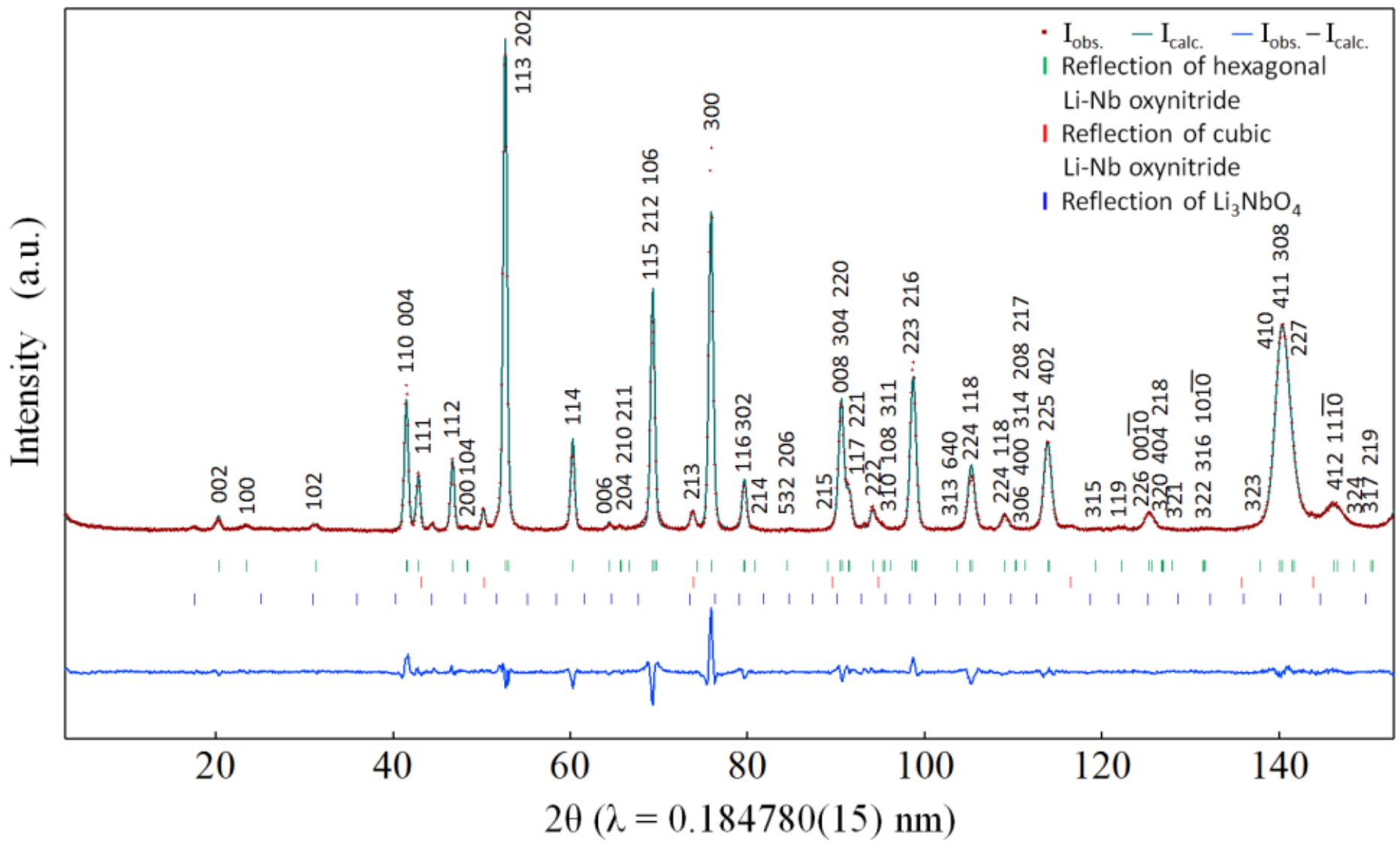

Fig. 3. Motohashi *et al.*

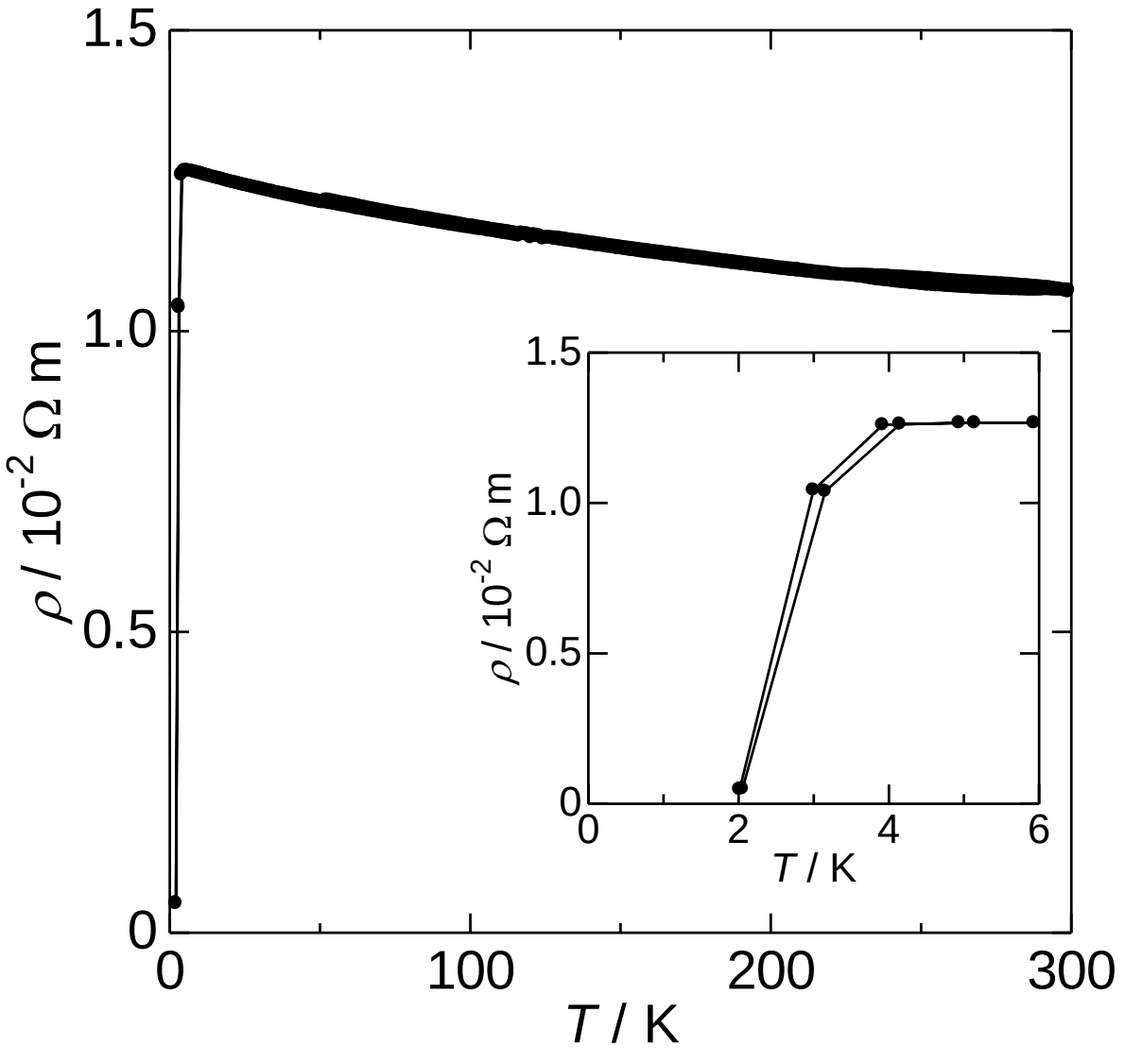

Fig. 4. Motohashi *et al*.

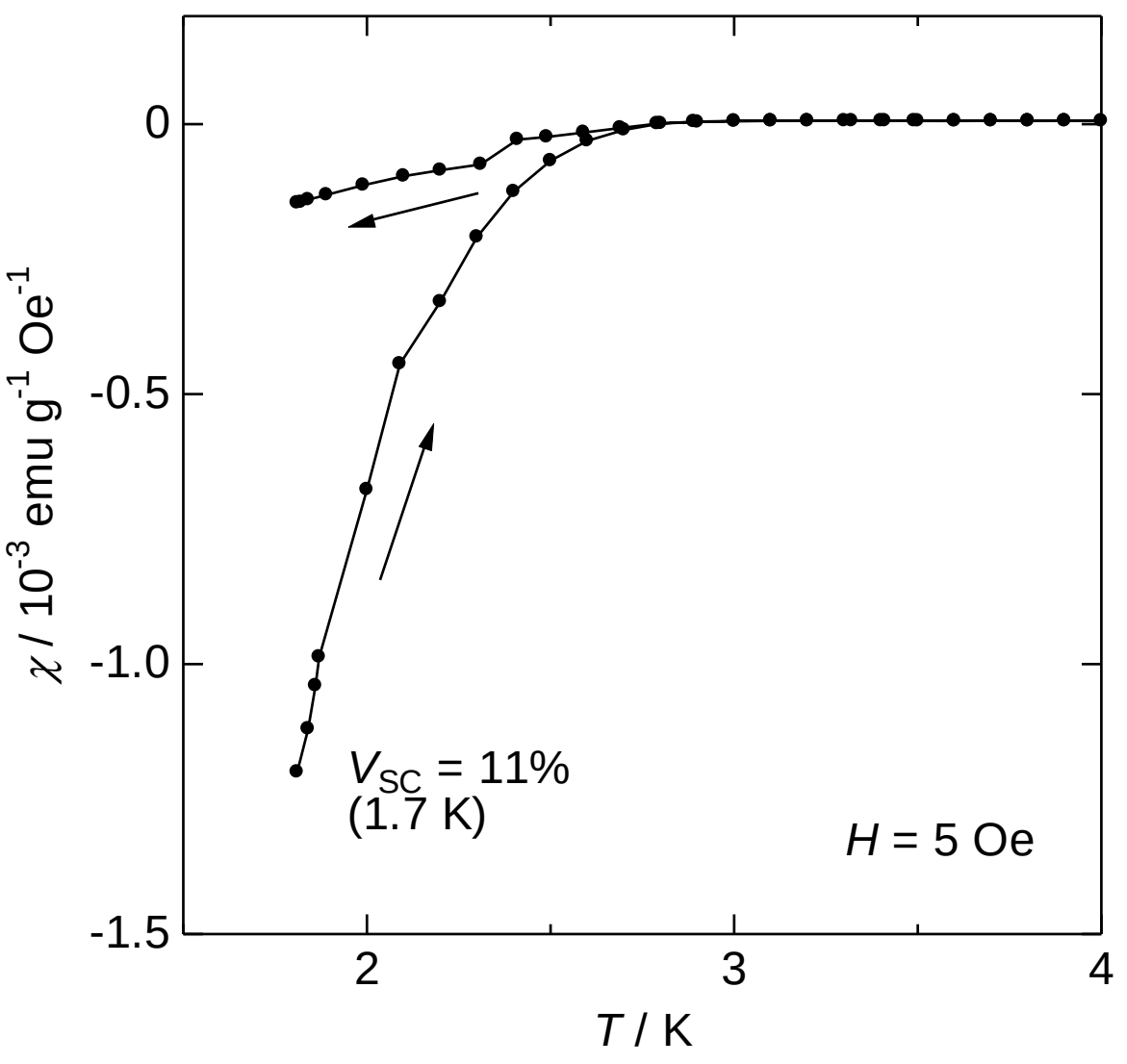

Fig. 5. Motohashi *et al.*

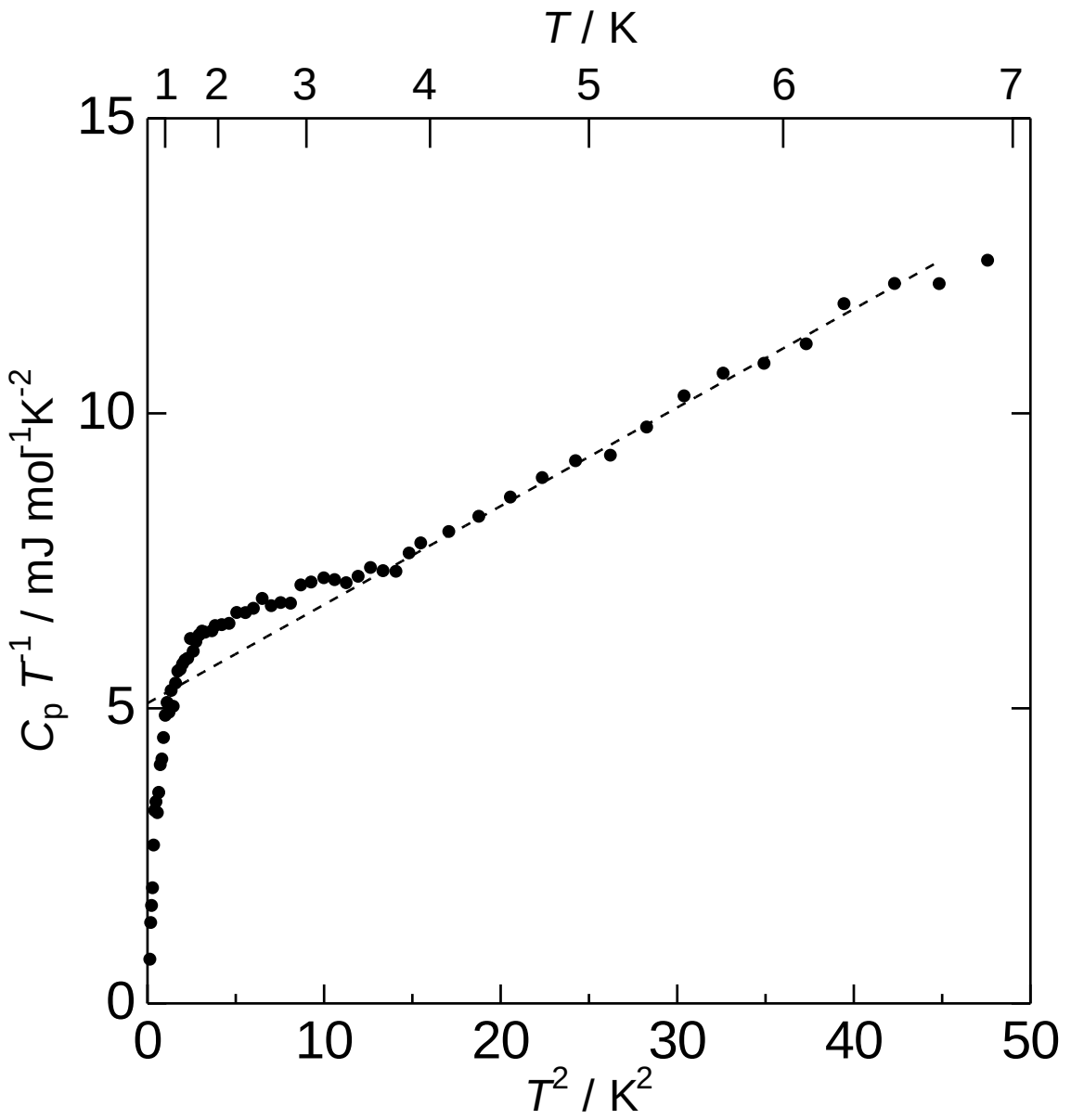

Fig. 6. Motohashi *et al*.

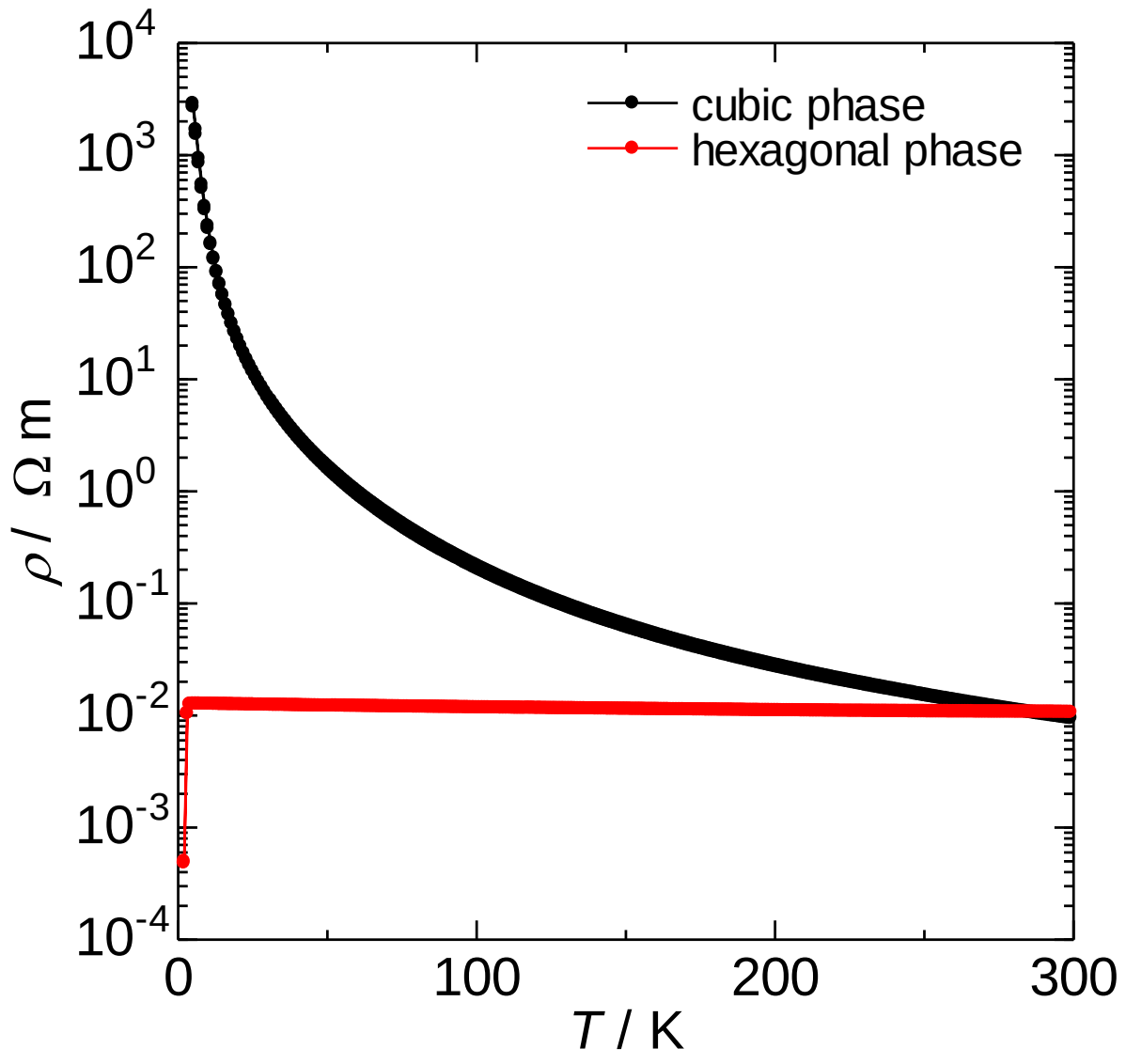

Fig. 7. Motohashi *et al.*